\documentclass[conference]{IEEEtran}
\usepackage{amsmath}
\usepackage{amssymb}
\usepackage{amsfonts}
\usepackage{amsthm}
\usepackage{algorithm}
\usepackage{algorithmic}
\usepackage{graphicx}
\usepackage{array}

\usepackage[shortlabels]{enumitem}
\usepackage{cite}
\usepackage{color}
\usepackage{epsfig,epstopdf}
\usepackage{url}
\usepackage{subfigure}
\usepackage{balance}
\usepackage{multirow}
\usepackage{verbatim} 
\usepackage[letterpaper,left=0.68in,right=0.68in,top=0.7in,bottom=1.05in]{geometry}

\IEEEoverridecommandlockouts

\begin{document}

\title{Sensor-Based Satellite IoT for Early Wildfire Detection}

\author{\IEEEauthorblockN{How-Hang Liu{\textsuperscript{$1$}}, Ronald Y. Chang{\textsuperscript{$1$}}, Yi-Ying Chen{\textsuperscript{$2$}}, and I-Kang Fu{\textsuperscript{$3$}}}
\IEEEauthorblockA{
{\textsuperscript{$1$}}Research Center for Information Technology Innovation, Academia Sinica, Taiwan \\
{\textsuperscript{$2$}}Research Center for Environmental Changes, Academia Sinica, Taiwan \\
{\textsuperscript{$3$}}MediaTek Inc. \\
}
\IEEEauthorblockA{Email: \{liuhowhang, rchang\}@citi.sinica.edu.tw, yiyingchen@gate.sinica.edu.tw, ik.fu@mediatek.com}
\thanks{This work was supported in part by the Ministry of Science and Technology, Taiwan, under Grant MOST 109-2221-E-001-013-MY3.}}

\maketitle

\begin{abstract}
Frequent and severe wildfires have been observed lately on a global scale. Wildfires not only threaten lives and properties, but also pose negative environmental impacts that transcend national boundaries (e.g., greenhouse gas emission and global warming). Thus, early wildfire detection with timely feedback is much needed. We propose to use the emerging beyond fifth-generation (B5G) and sixth-generation (6G) satellite Internet of Things (IoT) communication technology to enable massive sensor deployment for wildfire detection. We propose wildfire and carbon emission models that take into account real environmental data including wind speed, soil wetness, and biomass, to simulate the fire spreading process and quantify the fire burning areas, carbon emissions, and economical benefits of the proposed system against the backdrop of recent California wildfires. We also conduct a satellite IoT feasibility check by analyzing the satellite link budget. Future research directions to further illustrate the promise of the proposed system are discussed.
\end{abstract}

\IEEEpeerreviewmaketitle

\section{Introduction}\label{sec:intro}

Global warming is a pressing global issue and a result of excessive green house gas (GHG) emission. A major element of GHG is carbon dioxide. There are many contributing factors to carbon emissions, including wildfires. Global carbon emissions from fires for the recent years are $2.0$ GtC yr$^{-1}$ which is $40\%$ of the atmospheric carbon dioxide growth rate \cite{ORCHIDEE}. Frequent and intense fire activities could switch forests from carbon stocks to sources, especially during drought times, and such extreme cases contribute to significant carbon emission. A timely wildfire detection system can reduce carbon emissions and negative environmental impacts, as well as save lives and properties, by shifting the fire regime to a short and small one through reducing the spread of fires.

Satellite-based monitoring has been used in detecting wildfires. However, the long scan period and low resolution of satellites limited the performance of the system \cite{RTFF}. Moreover, the two-dimensional image data from satellite-sensor remote sensing cannot present the landscape-level ecological information such as biomass, soil wetness, and wind speed, which results in inaccurate detection \cite{RST}. These challenges inspired the employment of sensor-based systems \cite{RAI}. Although the sensor-based systems could report more timely and precise information, there remain issues to address, such as deployment robustness, sensor power management, and communication reliability and security \cite{WSN_FEA}.

Various types of sensor-based systems such as observation by watchtowers and unmanned aerial vehicles (UAVs) \cite{IWSNFWD,IMF,UAVIOT} all face some of these issues. The installation and maintenance of watchtowers are costly \cite{IWSNFWD,IMF}. Some conventional watchtowers required human operation which added to the total cost of watchtowers. Therefore, Ko {\it et al.}\cite{IWSNFWD} designed a computer vision system to replace human presence in watchtowers. Flexibility is another issue because adding or removing watchtowers will create redundancy or insufficiency in the observation coverage. Zhang {\it et al.} \cite{IMF} proposed algorithms to optimize the locations of watchtower deployment. 

Using UAVs for disaster management faces two major challenges in regard to power management and security\cite{UAVIOT}. First, UAVs are battery-powered and the recharge time could create a lapse in communicating essential data. Second, the observed data by UAVs might contain sensitive or private information, and therefore the use of UAVs is typically subject to government regulations regarding their sizes, ranges, and maximum flight heights.

There is a potential to leverage the advances of wireless communication and Internet of Things (IoT) technologies for geo-hazards prevention \cite{8895751}. To achieve timely wildfire detection and overcome the aforementioned limitations of watchtowers and UAVs, we propose to use the emerging beyond fifth-generation (B5G) and sixth-generation (6G) satellite IoT communication technology to enable massive sensor deployment in remote areas. Satellite communication has the advantage of covering an expansive area but is traditionally a closed-loop small ecosystem that requires dedicated, expensive satellite communication equipment (e.g., satellite phones), and is not available for large-scale commercial use. We propose to use the same radio interface and the same terrestrial radio communication technologies (i.e., B5G or 6G) with low-cost, consumer-grade wireless equipment for satellite IoT communication, which was shown feasible in \cite{DNBIOT}, for early wildfire detection. The main contributions of this paper include:
\begin{itemize}
\item We develop a mathematical model and perform the feasibility analysis of sensor-based satellite IoT for wildfire monitoring and detection. We propose a wildfire evolution model that takes into account wind speed and soil wetness. We propose a carbon emission model. We perform a satellite IoT feasibility check by calculating the satellite system capacity.
\item We use real environmental data such as wind, soil wetness, and biomass information to conduct simulations using California as a case study. We demonstrate the effect of sensor density on the fire burning time and the resulting carbon emissions and monetary costs. Benefits and implications of the proposed method are discussed.
\end{itemize}

\section{The Proposed System} \label{sec:PA}

\begin{figure}[t]
\begin{center}
\includegraphics[width=0.98\columnwidth]{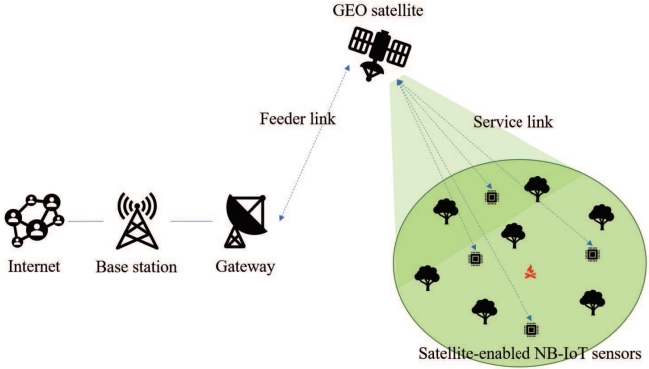}
\end{center}
\caption{The proposed satellite-enabled IoT sensors for wildfire detection.}
\label{fig:systemmodel}
\vspace{-0.1in}
\end{figure}

Fig.~\ref{fig:systemmodel} illustrates the proposed system. A massive number of geographically distributed sensors are deployed in remote areas (e.g., forests) for fire detection. The sensors are under the geostationary earth orbit (GEO) satellite beam coverage. Upon detecting a fire event, the sensors send alarm signals to the GEO satellite en route to the fire monitoring center for intervention and fire extinguishing measures.

%%%%%%%%%%%%%%%%%%%%%%%%%%%%%%%%%%%%%%%%%%%%%%%%%%%%%%%%%%%%%%%%%
%%%%%%%%%%%%%%%%%%%%%%%%%%%%%%%%%%%%%%%%%%%%%%%%%%%%%%%%%%%%%%%%%

We propose mathematical models to quantitatively evaluate the proposed system. In what follows, we first model the wildfire evolution. Then, we model the carbon emission according to the burned area and the biomass within the area. Finally, we calculate the satellite communication link budget to verify the number of sensors that can be supported by the system.

\subsection{Wildfire Model}

We extend the elliptical wildfire model with a single ignition point and a constant wind magnitude and direction \cite{FireSpread}, to wildfire evolution that involves multiple ignition points and varying wind magnitudes and directions as time evolves. These are described as follows.

\subsubsection{Elliptical Fire Model}

%%%%%%%%%%%%%%%%%%%%%%%%%%%%%%%%%%%%%%%%%%%%%%%%%%%%%%%%%%%%%%%%%
\begin{figure}[t]
\begin{center}
\subfigure[]{
    \label{fig:fire_ellipse}
    \includegraphics[width=0.45\columnwidth]{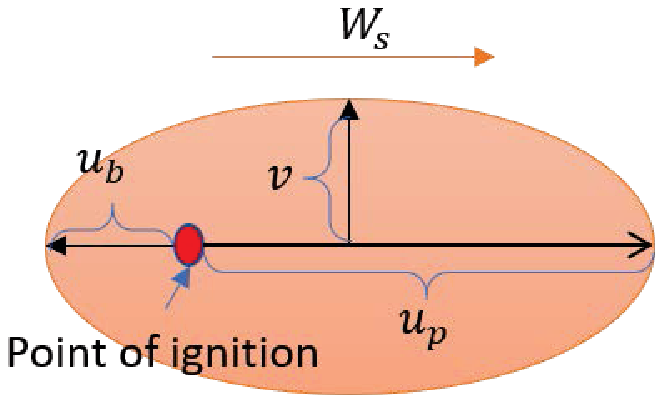}}
\subfigure[]{
    \label{fig:4point}
    \includegraphics[width=0.65\columnwidth]{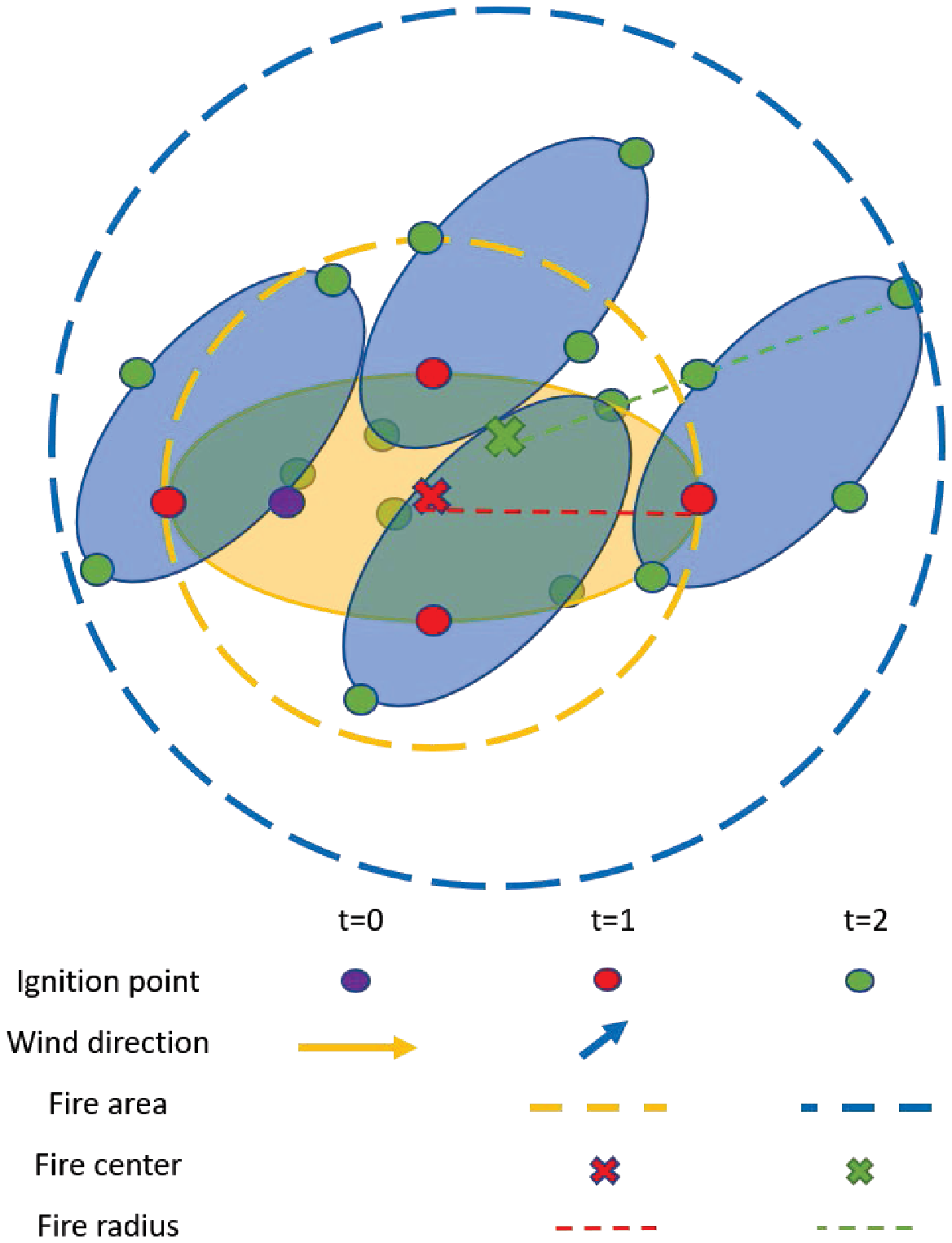}}
\caption{(a) Elliptical fire model. (b) Fire evolution model.}
\label{fig:fire_model}
\end{center}
\vspace{-0.1in}
\end{figure}
%%%%%%%%%%%%%%%%%%%%%%%%%%%%%%%%%%%%%%%%%%%%%%%%%%%%%%%%%%%%%%%%%

The spread of fire from a single ignition point under constant wind condition can be modeled as an ellipse, as depicted in Fig.~\ref{fig:fire_ellipse}. $W_s$ is the wind speed; $u_p$ is the fire spreading speed in the same direction of the wind (i.e., in the major axis direction of the ellipse); $u_b$ is the fire spreading speed in the opposite direction of the wind, where $u_b=0.2u_p$; and $v$ is the fire spreading speed in the minor axis direction of the ellipse. The fire spreading speed $u_p$ (m/s) can be represented as  
\begin{equation} \label{eq:u_p}
u_p=u_{\rm max}\times g(W_s)\times h(\beta_{\rm root}) 
\end{equation}
where $u_{\max}$ is a constant representing the maximum fire spreading speed which is $0.13$ m/s ($0.45$ km/h), and $g(W_s)$ and $h(\beta_{\rm root})$ are functions representing the dependence of $u_p$ on the wind speed ($W_s$) and root zone soil wetness ($\beta_{\rm root}$). The outputs of these two functions have values between $0$ and $1$. $g(W_s)$ is computed by
\begin{equation} \label{eq:gW_s}
g(W_s)=1-(1-g_0)\exp\left(-\frac{W_s^2}{2500}\right)
\end{equation}
where $g_0=0.1$ is a constant that controls the growth rate when there is no wind. $h(\beta_{\rm root})$ is described as 
\begin{equation} \label{eq:hbetaroot}
h(\beta_{\rm root})=(1-\beta_m)^2, \,\, \beta_m=
\begin{cases}
\frac{\beta_{\rm root}}{\beta_e}, & $for$\: \beta_{\rm root}\le\beta_e\\ 
1, & $for$\: \beta_{\rm root}> \beta_e
\end{cases}
\end{equation}
where $\beta_e=0.35$ is a threshold. When the soil wetness is above this threshold, $h(\beta_{\rm root})=0$, which leads to $u_p=0$. The $v$ speed can be calculated from the ellipse length-to-breadth ratio $L_B$ = $(u_p+u_b)/2v$. $L_B$ is related to $W_s$ and is represented as 
\begin{equation} \label{eq:LBratio}
L_B=1+10\big(1-\exp(-0.017W_s)\big).
\end{equation}
Since $u_p$, $u_b$, and $v$ are speeds, they will be multiplied by a time parameter to represent the distances of major and minor axes of the ellipse.

\subsubsection{Fire Evolution Model}

We model the fire development with changing winds as time evolves based on the elliptical fire model. Fig.~\ref{fig:4point} describes the proposed fire evolution model. At time $t=0$, there is a single ignition point and a single ellipse with the wind direction depicted. As time evolves, theoretically, each point inside the elliptical area for $t=0$ could be the next ignition point for $t=1$. This entails infinitely many points. For modeling tractability, we consider only the four points at the two ends of major and minor axes of each ellipse as the next ignition points, to approximate fire spreading. In other words, each ellipse at time $t$ will generate four ellipses at time $t+1$. As shown in Fig.~\ref{fig:4point}, the winds may change magnitudes and directions, resulting in different shapes and rotations of ellipses, as time evolves.

We approximate the fire burning area covered by the many ellipses by the area of a circle, since directly calculating the total area of possibly overlapping ellipses is intractable. The center of the circle for calculating the fire burning area at time $t$ is determined by the average coordinates of the next ignition points at time $t$, and the radius of the circle is determined by the longest distance from the center to any of the ignition points. For example, at time $t=2$, there are four shaded ellipses and $16$ next ignition points, as shown in Fig.~\ref{fig:4point}. The center of the circle is the average coordinates of the $16$ ignition points, and the radius of the circle is the longest distance from the center to any of the $16$ ignition points. The area of the circle, depicted by the blue dashed circle in Fig.~\ref{fig:4point}, will be used to approximate the fire burning area at $t=2$. While this may appear to be an overestimate of area, it is a reasonable approximation considering the fact that in reality there are many ellipses (instead of four) initiated from many ignition points, as mentioned previously. Note that while at $t=1$ we can calculate the fire burning area exactly (since there is only one ellipse), we still adopt the same approximation method for consistency.

\subsection{Carbon Emission Model} \label{CEM}

Given the burned area, and the biomass within the area, the corresponding carbon emission can be derived. Since the biomass varies from place to place, we use the average biomass to calculate the amount of carbon emissions. It is known that the underground biomass is $20\%$ of above-ground biomass \cite{RBA}, and both underground and above-ground biomass will contribute to carbon emission. Let the total burned area be $A$ (km$^2$) and the average (above-ground) biomass be $B_{\rm avg}$ (Mg/ha), where Mg/ha denotes megagram (or ton) per hectare (or $10^4$ m$^2$). Then, the carbon emission in the unit of ton is given by 
\begin{equation} \label{eq:carbon_emission}
A \times 1.2 B_{\rm avg}\times 100
\end{equation}
where $100$ is a unit conversion factor.

\begin{figure}[t]
\begin{center}
\includegraphics[width=0.35\columnwidth]{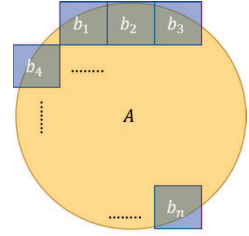}
\end{center}
\caption{Calculation of the average biomass for the carbon emission model.}
\label{fig:carbon_biomass}
\vspace{-0.1in}
\end{figure}

We calculate the average biomass by dividing the fire burning area into several (say, $N$) smaller homogeneous areas with biomass $b_n, n=1,\ldots,N$, and averaging them to get the average biomass $B_{\rm avg}=(b_1+b_2+\cdots+b_N)/N$ for the fire burning area, as illustrated in Fig.~\ref{fig:carbon_biomass}. In our model, each small area has an area of one hectare with a biomass value.

%%%%%%%%%%%%%%%%%%%%%%%%%%%%%%%%%%%%%%%%%%%%%%%%%%%%%%%%%%%%%%%%%
%%%%%%%%%%%%%%%%%%%%%%%%%%%%%%%%%%%%%%%%%%%%%%%%%%%%%%%%%%%%%%%%

\subsection{Satellite Link Budget Analysis}

Here, we conduct a feasibility check of GEO satellite communication supporting sensors on the earth based on the 3GPP NB-IoT non-terrestrial network (NTN) solution \cite{systemOverview}. We assume that the sensors are uniformly distributed in the sensing area, and the sensing area is within the satellite beam coverage. The GEO satellite beam diameter is $1000$ km. Table~\ref{tab:linkanalysis} summarizes the GEO satellite communication parameters for two cases of elevation angle \cite{systemOverview}. $10$-degree elevation angle represents the worst case of link budget due to the largest distance between the satellite and the sensors, whereas $90$-degree elevation angle represents the best case of link budget as the satellite is directly overhead. Other cases of link budget fall between these two cases.

%%%%%%%%%%%%%%%%%%%%%%%%%%%%%%%%%%%%%%%%%%%%%%%%%%%%%%%%%%%%%%%%%
	\begin{table}[t]
	\begin{center}
	\caption{GEO Satellite Link Budget Parameters}
	\label{tab:linkanalysis} \vspace*{0in}
	\begin{tabular}{|c||c|c|}
	\hline
	Elevation angle (degrees) & $10$ & $90$ \\ \hline
	Transmission mode & UL & UL \\ \hline
	Subcarrier frequency (GHz) & $1.5$ & $1.5$\\ \hline
    TX: EIRP (dBm) & $23$ & $23$\\ \hline
    RX:	G/T (dB/K) & $19$ & $19$\\ \hline
    Bandwidth (kHz) & $3.75$ & $3.75$ \\ \hline
    ${\rm PL_{FS}}$ (dB) & $188.14$ & $187.05$ \\ \hline
    ${\rm PL_{A}}$ (dB) & $0.16$ & $0.16$\\ \hline
    ${\rm PL_{SM}}$ (dB) & $3$ & $3$\\ \hline
    ${\rm PL_{S}}$ (dB) & $2.2$ & $2.2$\\ \hline
    ${\rm PL_{P}}$ (dB) & $3$ & $3$\\ \hline
    Distance (km) & $40581$ &$35786$ \\ \hline
    CNR (dB) &$8.3714$ & $9.4636$\\ \hline
	\end{tabular}
	\end{center}
	\vspace{-0.1in}
	\end{table}
%%%%%%%%%%%%%%%%%%%%%%%%%%%%%%%%%%%%%%%%%%%%%%%%%%%%%%%%%%%%%%%%%

The received carrier-to-noise ratio (CNR) at the satellite for the uplink transmission can be calculated by \cite{38821}
\begin{align}
{\rm CNR\:(dB)} &= {\rm EIRP\:(dBW)}+{\rm G/T\:(dB/K)} \nonumber\\
&\quad -{\rm PL_{FS}\:(dB)}-{\rm PL_{A}\:(dB)}-{\rm PL_{SM}\:(dB)} \nonumber\\
&\quad -{\rm PL_{S}\:(dB)}-\:{\rm PL_{P}\:(dB)} \nonumber\\
&\quad -{\rm BW\:(dBHz)}-k\:{\rm (dBW/K/Hz)} \label{eq:SNR}
\end{align}
where all terms are specified in Table~\ref{tab:linkanalysis} except the last term which is the Boltzmann constant $k=-228.6$. The first term of \eqref{eq:SNR} is the effective isotropic radiated power (EIRP) equal to $23$ dBm. The second term of \eqref{eq:SNR} is the antenna-gain-to-noise-temperature ratio (G/T) equal to $19$ dB/K. The third term of \eqref{eq:SNR} is the free space path loss (${\rm PL_{FS}}$) equal to $188.14$ dB when the elevation angle is $10$ degrees, which is calculated from
\begin{equation} \label{eq:FSPL}
{\rm PL_{FS}} = 32.45\:({\rm dB})+20\log_{10}(d)+20\log_{10}(f)
\end{equation}
where $f=1.5$ GHz and $d=40581$ km from Table~\ref{tab:linkanalysis}. The fourth term of \eqref{eq:SNR} is the atmospheric path loss (${\rm PL_{A}}$) due to gases and rain fades, taken to be $0.16$ dB. The fifth term of \eqref{eq:SNR} is the shadow fading margin (${\rm PL_{SM}}$) due to obstacles affecting the wave propagation, taken to be $3$ dB. The sixth term of \eqref{eq:SNR} is the scintillation loss (${\rm PL_{S}}$) caused by the local variation of the ionospheric electron density, taken to be $2.2$ dB. The seventh term of \eqref{eq:SNR} is the polarization loss (${\rm PL_{P}}$) caused by the polarization mismatch in antennas, taken to be $3$ dB. The eighth term of \eqref{eq:SNR} is the channel bandwidth (BW) which is $3.75$ kHz. Summing up these terms gives $\mbox{CNR}=8.3714$ dB for the worst case ($10$-degree elevation angle). This CNR value, according to the transport block size (TBS) table in the 3GPP standard \cite{25214}, allows the sensors to transmit $144$ bits per resource unit for uplink single tone (subcarrier) transmission. A resource unit is $32$ ms in time and $3.75$ kHz in frequency. Since the total available system bandwidth is $180$ kHz, the peak data rate (or throughput) is $144\times(180/3.75/32)$ bits/ms or $216$ kbps for the worst case ($10$-degree elevation angle).

We consider two typical cases of wildfire sensing based on the sensor traffic model in \cite{naroIOT}: periodic report and fire-event-triggered report. For the first case, the sensor report rate is twice per day and the sensing data size per connection is $50$ bytes, which leads to $0.0093$ bps under the assumption of uniform traffic distribution. For the second case, the sensor report rate is once per minute with the same data size, which leads to $6.6667$ bps. The supportable numbers of sensors operating in the first and second cases can be calculated as $216 \mbox{ (kbps)}/0.0093 \mbox{ (bps)} = 2.32\times 10^7$ and $216 \mbox{ (kbps)}/6.6667 \mbox{ (bps)} = 3.24\times 10^4$, respectively. Considering deploying one sensor per acre, or about $247$ sensors per km$^2$, the supportable number of sensors for periodic report can cover most of the forest area in the U.S. \cite{SLNF}. On the other hand, since a fire event will trigger only sensors nearby at a time as fire propagates, a capacity to support $3.24\times 10^4$ sensors for simultaneous event-triggered report is sufficient for practical use. Clearly, increasing the system bandwidth (which incurs additional operational cost) will increase the link budget and the supportable number of sensors for both cases of wildfire sensing. 

\section{Simulation Results} \label{sec:simulation}

\subsection{Simulation Setup: California Wildfires}

\begin{figure}[t]
\begin{center}
\includegraphics[width=\columnwidth]{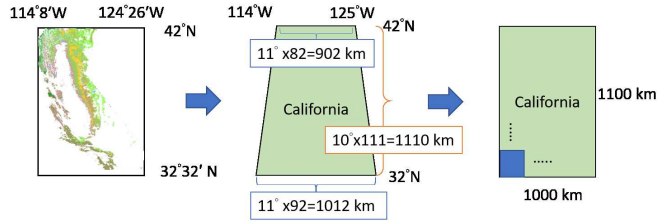}
\end{center}
\caption{Simulation area approximation of California.}
\label{fig:fire_area}
\vspace{-0.1in}
\end{figure}

We consider California wildfires with fundamental historical wildfire data as a case study for our proposed model. Historically, California has the highest number of wildfires and burned areas among all states in the U.S. \cite{FSW}, motivating a meaningful case study. The latitude and longitude of California are $32^{\circ}~32'$ N to $42^{\circ}$ N and $114^{\circ}~8'$ W to $124^{\circ}~26'$ W, respectively. Since $1^{\circ}$ of latitude equals approximately $111$ km, California spans about $1110$ km vertically. Since the distance between longitudes varies depending on the latitude, California spans about $1012$ km at $32^{\circ}$ N latitude and $902$ km at $42^{\circ}$ N latitude horizontally. For simplicity, we approximate the geographic area of California by a $1000\times 1100$ km$^2$ area, as illustrated in Fig.~\ref{fig:fire_area}, with grid cells corresponding to the environmental data described next.

%%%%%%%%%%%%%%%%%%%%%%%%%%%%%%%%%%%%%%%%%%%%%%%%%%%%%%%%%%%%%%%%%
%%%%%%%%%%%%%%%%%%%%%%%%%%%%%%%%%%%%%%%%%%%%%%%%%%%%%%%%%%%%%%%%

Table~\ref{tab:envirdata} presents the environmental data of California in 2020 from \cite{CDS}. There are four variables. $u_{10}$ is the $10$-meter U-wind component and $v_{10}$ is the $10$-meter V-wind component. The wind speed $W_s$ in our wildfire model can be calculated directly from $W_s=\sqrt{u_{10}^2+v_{10}^2}$, and the wind direction, represented by the angle $\theta$ with the horizontal axis, can be calculated as $\theta = \arctan(v_{10}/u_{10})$. Swvl1 is volumetric soil water in the unit of $\%$, given by the volume of water (m$^3$) divided by the volume of soil (m$^3$), which corresponds to $\beta_{\rm root}$ in \eqref{eq:u_p}. These three variables are each three-dimensional corresponding to longitude, latitude, and time. Specifically, the $1000\times 1100$ km$^2$ area depicted in Fig.~\ref{fig:fire_area} is divided into $111\times 101$ grid cells, and the time is in the unit of hours for the entire year of 2020, given by $366$ (day) $\times$ $24$ (hr/day) $= 8784$ (hr). The last variable is Biomass which is the amount of the above-ground live biomass in the unit of Mg/ha. The value of this variable is used to calculate the average biomass in our model in Sec.~\ref{CEM} to get the carbon emission. Biomass does not vary from time to time in our dataset but changes from location to location.

%%%%%%%%%%%%%%%%%%%%%%%%%%%%%%%%%%%%%%%%%%%%%%%%%%%%%%%%%%%%%%%%%
	\begin{table*}[t]
	\begin{center}
	\caption{The Environmental Data of California in 2020}
	\label{tab:envirdata} \vspace*{0in}
	\begin{tabular}{|c|c|c|c|c|}
	\hline
	Variables & $u_{10}$ & $v_{10}$ & Swvl1 & Biomass\\ \hline\hline
	Name & 10-meter U-wind component & 10-meter V-wind component & Volumetric soil water layer 1 & Above-ground live biomass \\ \hline
    Dimensions & $111\times101\times8784$ & $111\times101\times8784$ & $111\times101\times8784$ & $11645\times10666\times1$ \\ \hline
    Grid spacing & $10$ (km) & $10$ (km) & $10$ (km) & $100$ (m) \\ \hline
    Unit & m/s & m/s & \% & Mg/ha\\ \hline
	\end{tabular}
	\end{center}
	\vspace{-0.2in}
	\end{table*}
%%%%%%%%%%%%%%%%%%%%%%%%%%%%%%%%%%%%%%%%%%%%%%%%%%%%%%%%%%%%%%%%%

We compare our simulation results with the historical wildfire data from the California Department of Forestry and Fire Protection (CAL FIRE) database \cite{CALFIRE}. The database records the incident created time, incident extinguished time, incident coordination, and incident burned area in acre for all $255$ California wildfires in the entire year of 2020. We set the ignition point and starting time of each fire outbreak in our simulation according to these $255$ wildfires for a fair annual comparison with the historical data. The sensors are randomly deployed (uniformly distributed) in the approximate geographic area of California.

\subsection{Results and Discussion} \label{sec:Discussion}

\subsubsection{Fire Burning Time/Area vs. the Number of Sensors}

%%%%%%%%%%%%%%%%%%%%%%%%%%%%%%%%%%%%%%%%%%%%%%%%%%%%%%%%%%%%%%%%%
\begin{figure*}[t]
\begin{center}
\subfigure[]{
    \label{fig:Sec03_6a_BurnedHour}
    \includegraphics[width=0.65\columnwidth]{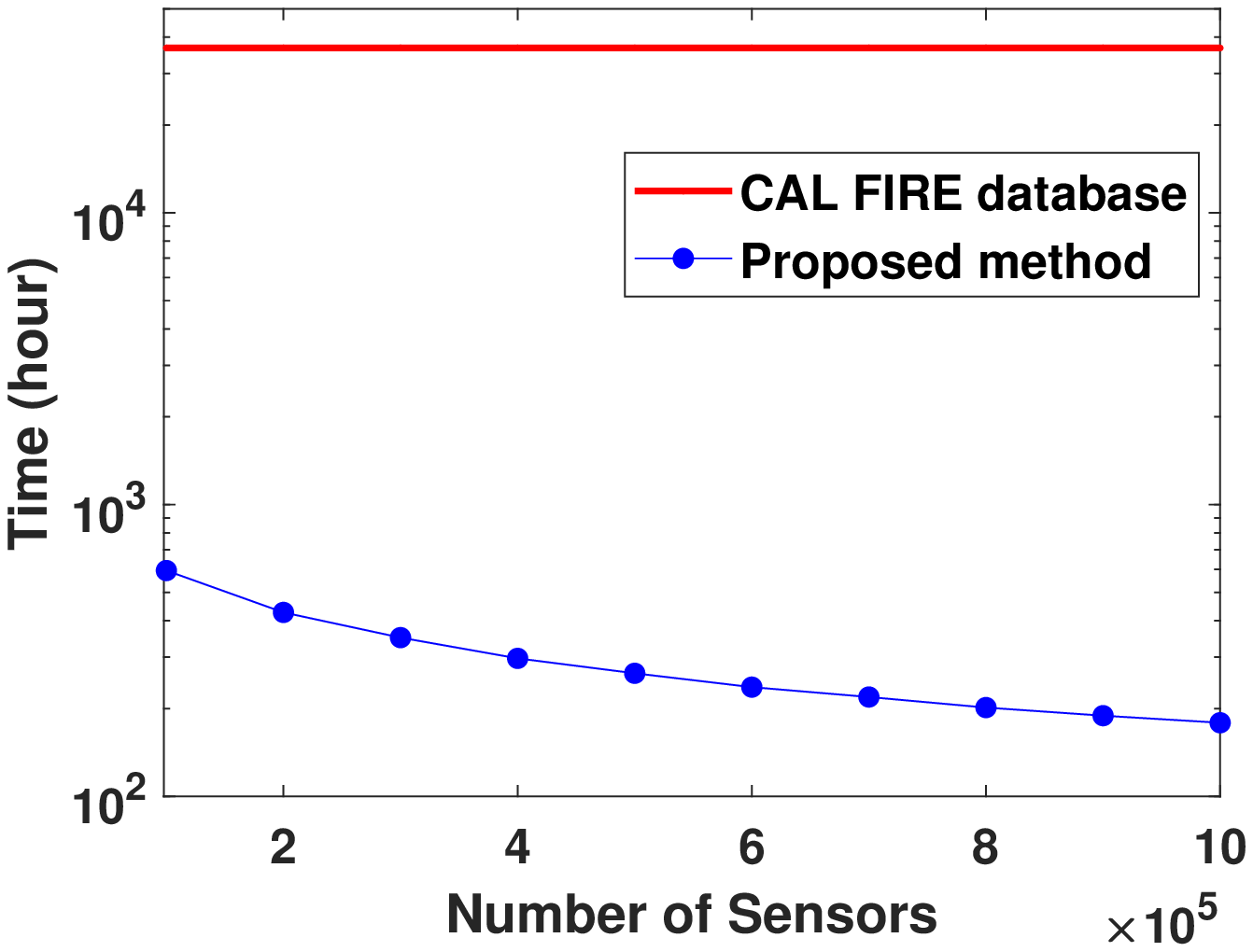}}
\subfigure[]{
    \label{fig:Sec03_6b_BurnedArea}
    \includegraphics[width=0.65\columnwidth]{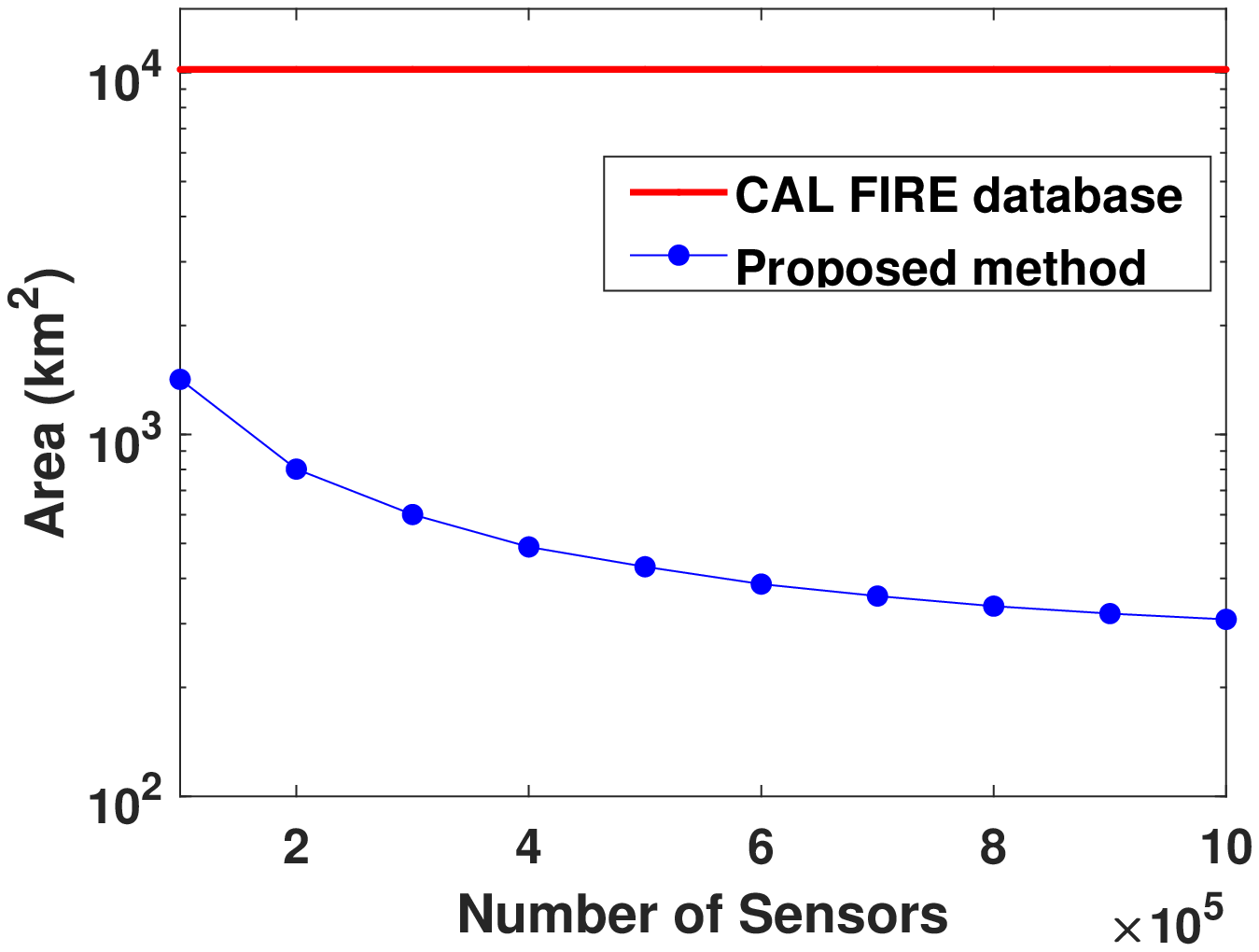}}
\subfigure[]{
    \label{fig:Sec03_6c_SensorCarbon}
    \includegraphics[width=0.65\columnwidth]{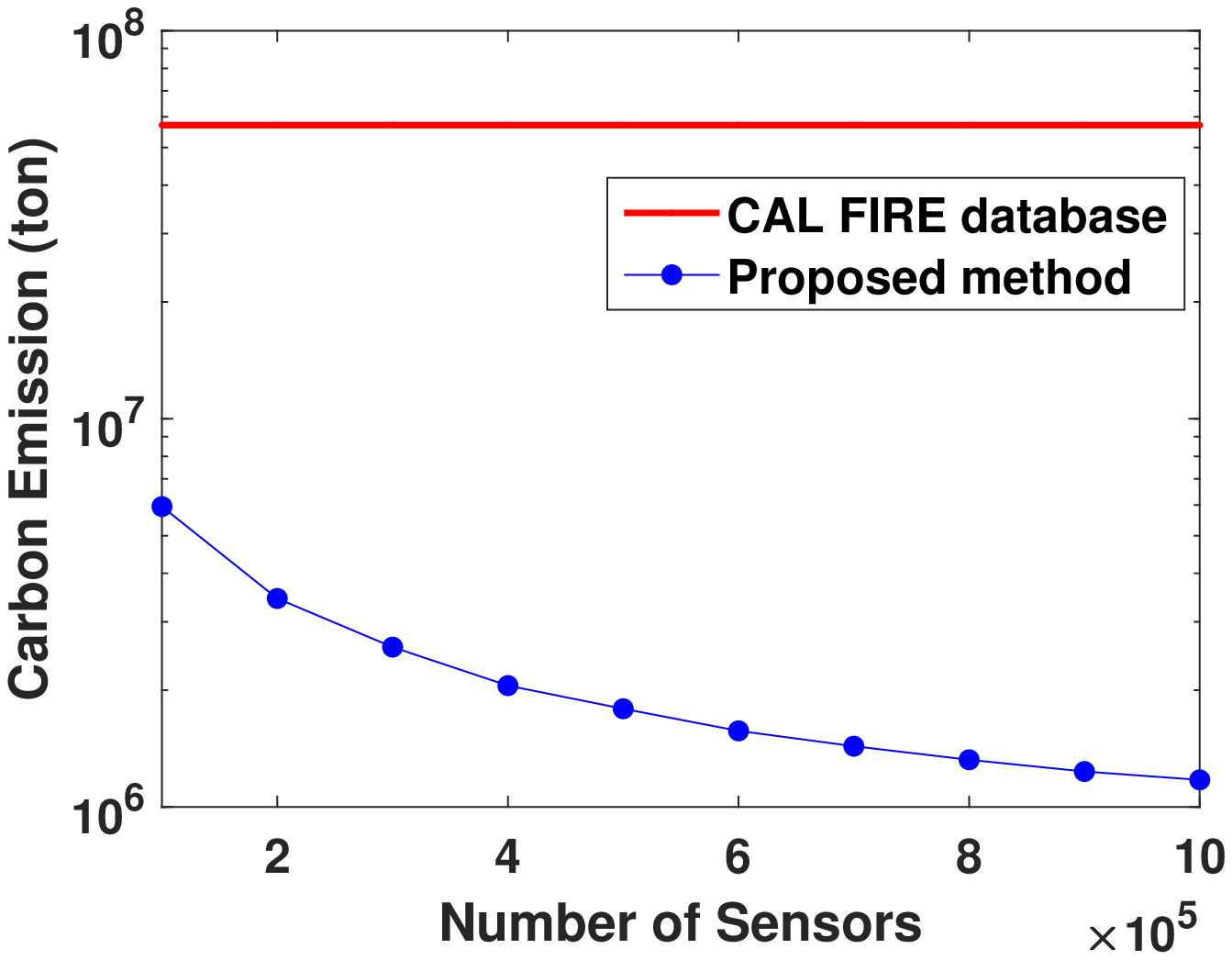}}
\subfigure[]{
    \label{fig:Sec03_6d_SensorPrice}
    \includegraphics[width=0.65\columnwidth]{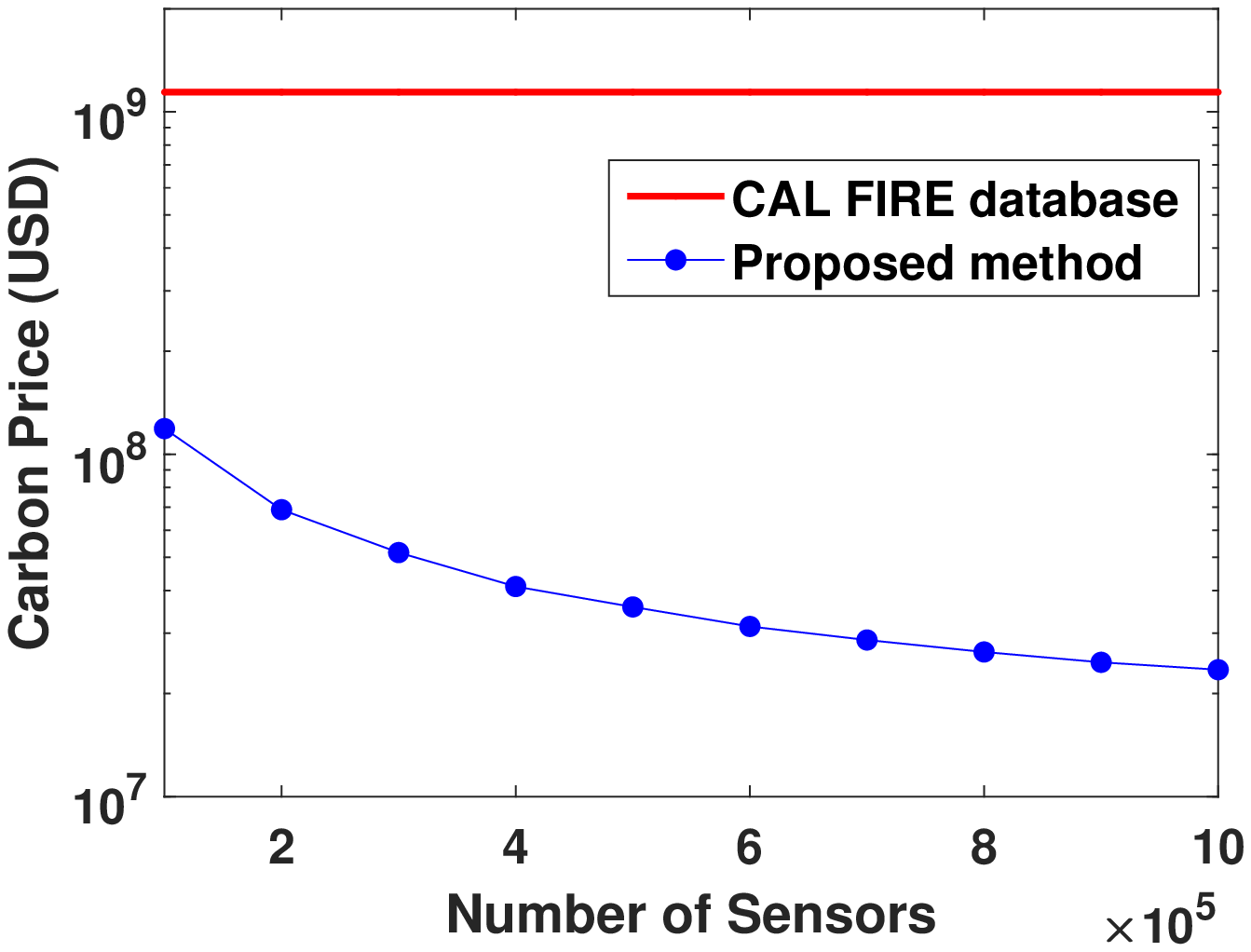}}
\subfigure[]{
    \label{fig:Sec03_6d_FinalProfit}
    \includegraphics[width=0.65\columnwidth]{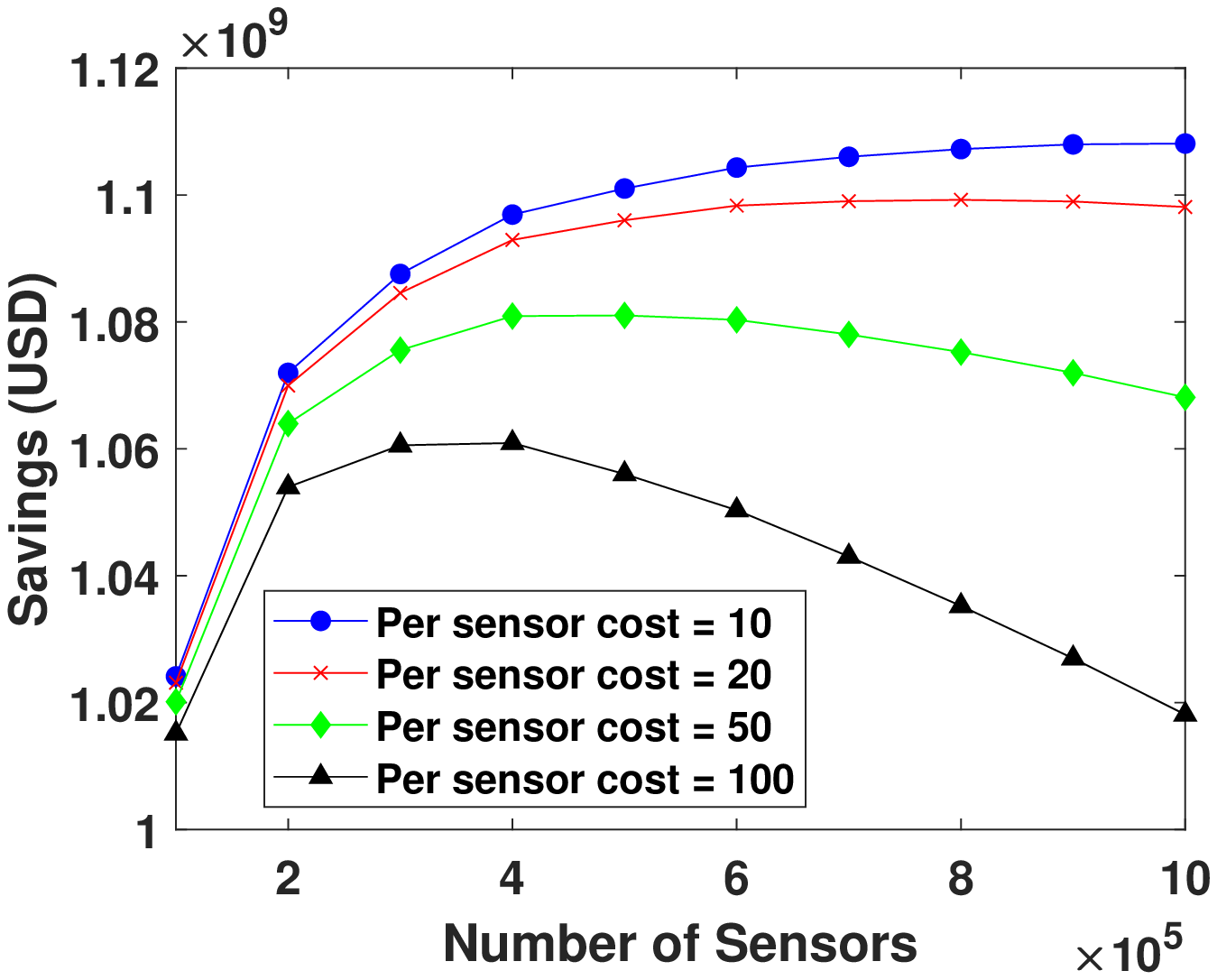}}
\caption{Simulation results. (a) Fire burning time (hours). (b) Fire burning area (km$^2$). (c) Carbon emission (ton). (d) Carbon price (USD). (e) Savings (USD), for sensor costs of $10$, $20$, $50$, and $100$ USD per sensor. The 2020 wildfire data from the CAL FIRE database are shown for comparison.}
\label{fig:simulation_result}
\end{center}
\vspace{-0.1in}
\end{figure*}
%%%%%%%%%%%%%%%%%%%%%%%%%%%%%%%%%%%%%%%%%%%%%%%%%%%%%%%%%%%%%%%%%

We first examine the effect of the number of sensors on the fire burning time and area. Fig.~\ref{fig:Sec03_6a_BurnedHour} shows the burned hours vs. the number of sensors result. For simplicity and to focus our discussion on the contribution of the number of sensors, we assume that the fire is extinguished as soon as the fire spreading reaches any of the sensors. As can be seen, deploying more sensors reduces the burned hours since the fire could be more timely detected. The number of sensors ranges from $10^5$ to $10^6$, where $10^6$ sensors deployed in the $1000\times 1100$ km$^2$ area of California amounts to approximately one sensor per km$^2$. The burned time when $10^5$ and $10^6$ sensors are deployed is $594.21$ hours and $178.77$ hours, respectively. In comparison, the historical annual burned time, calculated by summing the burning time (the duration between incident created time and incident extinguished time) of all California wildfires in 2020 in the CAL FIRE database, is $36716.25$ hours.

Fig.~\ref{fig:Sec03_6b_BurnedArea} shows the burned areas vs. the number of sensors result. The burned areas are $1420.11$ km$^2$ and $308.35$ km$^2$ when $10^5$ and $10^6$ sensors are deployed, respectively. In comparison, the annual burned area derived from the CAL FIRE database is $10202.04$ km$^2$.

\subsubsection{Carbon Emission/Price vs. the Number of Sensors}

We calculate the carbon emission based on the carbon emission model in \eqref{eq:carbon_emission}. Fig.~\ref{fig:Sec03_6c_SensorCarbon} shows the result. The amounts of carbon emission are $5.82\times10^6$ ton and $1.17\times10^6$ ton when $10^5$ and $10^6$ sensors are deployed, respectively. According to \cite{PCTofCA}, we assume per ton of carbon is equal to $20$ USD. The carbon price is shown in Fig.~\ref{fig:Sec03_6d_SensorPrice}, which is $116.4$ million USD with $10^5$ sensors and $23.4$ million USD with $10^6$ sensors.

In comparison, the annual carbon emission from the 2020 California wildfires can be calculated as $10202.04 \times (1.2 \times 46.6237) \times 100 = 5.71 \times 10^7$ ton based on \eqref{eq:carbon_emission}, where we have adopted the average biomass value of the entire California, $46.6237$ Mg/ha. This amounts to a carbon price of $1.14$ billion USD.

\subsubsection{Savings vs. the Number of Sensors}

The annual saving yielded by using our proposed sensor-based system is the monetary difference between the $1.14$ billion USD cost and the carbon price for the proposed method shown in Fig.~\ref{fig:Sec03_6d_SensorPrice} plus the sensor costs. For example, assuming the sensor cost of $100$ USD per sensor, the saving is $1.14\times 10^9 - (116.4\times 10^6+10^5\times 100)=1.01\times10^9$ USD, or $1.01$ billion USD, when $10^5$ sensors are deployed. The savings result is shown in Fig.~\ref{fig:Sec03_6d_FinalProfit} for four different sensor costs. As can be seen, the proposed sensor-based system can potentially lead to significant annual savings by detecting wildfires early.

\subsection{Further Discussion and Future Directions}

The results presented have shown the promise of the proposed system. Further studies in the following directions can be done to enhance the completeness and generality of the proposed system:
\begin{itemize}
\item The current study considers an ideal situation where fire is contained immediately upon detection of fire by one or more sensors. The time and cost associated with the follow-up fire extinguishing measures should be taken into account to provide a more well-rounded view of the total savings achieved by the proposed method.
\item The current study considers a simple sensor placement model (i.e., uniform distribution). In practice, the terrain (land, sea, lake, etc.), population density (urban/suburban areas, etc.), and heterogeneous characteristics and vegetation of the land coverage area may be considered for a more intelligent and efficient placement of sensors. For example, more sensors should be deployed in high-risk, high-flammability areas such as forests. This consideration can be further coupled with an analysis of the critical fire time and area beyond which a fire would be difficult to control, so as to identify the required number of sensors in specific areas.
\item The current study could further take into account different causes of fire, especially, natural causes of fires (e.g., lightning) and human-caused fires, which may demand different sensor types and data traffic models. \item A proof-of-concept field trial can be conducted to verify the integration and testing of wildfire detection sensors and NB-IoT NTNs.
\end{itemize}

\section{Conclusion} \label{sec:conclusion}

In this paper, we have proposed a sensor-based satellite IoT system for wildfire detection. We quantitatively investigated the feasibility and effectiveness of the proposed system. Specifically, we first proposed a wildfire model with multiple ignition points and varying wind magnitudes and directions, and a method to approximate the fire burning area. We then proposed a carbon emission model based on real biomass information, as well as conducted a satellite link budget analysis. Simulation results based on real environmental data of California in 2020 demonstrated that deploying as few as one sensor per km$^2$ could reduce the annual carbon emission by more than ten times, and deploying our system could yield significant annual savings of billions of USD due to early fire containment. Pointers on further research based on this pilot study were outlined.

\bibliographystyle{IEEEtran}
\bibliography{IEEEabrv,ref}
\balance

\end{document}